\documentclass[aps,pre,showpacs,twocolumn]{revtex4}
\usepackage{graphics}
\usepackage{psfrag}
\usepackage{graphicx}
\usepackage{times}

\begin{document}

\title{Similarity based cooperation and spatial segregation}
\author{Arne Traulsen} 
\email{traulsen@theo-physik.uni-kiel.de}
\author{Jens Christian Claussen}
\email{claussen@theo-physik.uni-kiel.de}
\affiliation{Institut f{\"u}r Theoretische Physik und
Astrophysik, Christian-Albrechts Universit{\"a}t,
Olshausenstra{\ss}e 40, 24098 Kiel, Germany}
\date{August 11, 2004; published 29 October 2004 in Physical Review E }

\begin{abstract}
We analyze a cooperative game, where the cooperative act is not based
on the previous behaviour of the coplayer, but on the similarity between
the players. This system has been studied in a mean-field description recently
[A.\ Traulsen and H.\ G.\ Schuster, Phys.\ Rev.\ E {\bf 68}, 046129 (2003)].
Here, the spatial extension to a two-dimensional lattice is studied, where
each player interacts with eight players in a Moore neighborhood.
The system shows a strong segregation independent of parameters.
The introduction of a local conversion mechanism towards tolerance
allows for four-state cycles and the emergence of spiral waves in the spatial game.
In the case of asymmetric costs of cooperation a rich variety of
complex behavior is observed depending on both cooperation costs.
Finally, we study the stabilization of a cooperative fixed point 
of a forecast rule in the
symmetric game, which corresponds to cooperation across segregation
borders. This fixed point becomes unstable for high cooperation costs,
but can be stabilized by a linear feedback mechanism.
\end{abstract}
\pacs{
02.50.Le,		
87.23.-n,		
89.65.-s		
}

\maketitle

\section{Introduction}

The study of complex systems with game 
theoretic interactions has recently attracted a lot
attention in statistical physics, biology,
behavioral sciences, and economics. In contrast to
standard game theory \cite{Neumann1953} the 
focus has recently been on evolutionary 
game theory \cite{Smith1982,Gintis2000,Hofbauer1998,Nowak2004a,Nowak2004b}.
In particular, the prisoner's dilemma \cite{Axelrod1984} 
has become the metaphor for the evolution of 
cooperation in populations of self\hspace*{0.001mm}ish individuals. 
If the game is not repeated, the dominating strategy is
to defect. However, repeated interactions
of individuals memorizing the past can establish
high levels of cooperation from direct reciprocity 
\cite{Axelrod1984}. Reciprocity works also
indirectly if individuals can observe the behavior
of others and cooperate with respect to the 
reputation of others \cite{Nowak1998,Nowak1998b}.

Real world interactions are often 
restricted to small local groups. 
Realizing that territoriality can have strong 
influences on the evolution of cooperation, 
Axelrod proposed the study of a spatially 
extended prisoner's dilemma \cite{Axelrod1984}. 
Nowak and May studied a cellular automaton based on
the prisoner's dilemma \cite{Nowak1992}. They found
that reducing interactions to small local groups 
can promote cooperation, as cooperative clusters 
minimize their contacts with neighboring defectors.
Their paper initiated an intensive research 
on spatially extended games
on two dimensional lattices
\cite{Lindgren1994,Herz1994,Szabo1998,Szabo2002}
and network topologies \cite{Abramson2001,Ebel2002,Holme2003,Szabo2004}.
However, spatial structure does not necessarily lead to an increased level of cooperation
\cite{Hauert2004}.

Another mechanism that can promote cooperation
among related individuals is kin selection 
\cite{Hamilton1963}.
Although kin selection is controversial in biology,
indications for similarity based interaction mechanisms 
have been found on the molecular level \cite{Turner1999,Taga2003}.
Riolo {\em et al.} introduced a model in 
which agents are equipped with traits that 
allow one to discriminate between different groups of
players \cite{Riolo2001}. It has been argued that
the model is of limited biological relevance,
as agents are forced to cooperate within their
own group \cite{Roberts2002, Hauert2002c}. 
However, 
cooperation can evolve from a combination of kin selection
and reciprocity, which can promote such an
intragroup cooperation. The mechanism that 
leads to cooperation does not have to be the same 
for interaction within groups and between groups.
In spatially extended systems agents can 
only prosper if they get suff\hspace*{0.001mm}icient
support from their neighbors. Hence, cooperation
based on similarity will lead to a segregation
of different groups in spatially extended system
\cite{Sigmund2001}.

Although the importance of group memberships is 
stressed in the social sciences \cite{Hammond2003}, 
segregation is usually not desired in social systems. 
We raise the question on the minimal requirements 
for agents in order to avoid this kind of segregation. 
We introduce a forecast rule that helps to overcome the
segregation, leading to a population in which agents
support others regardless of their group membership. 
The corresponding spatial pattern can be stabilized by
a linear global feedback. 

\section{Definition of the model}

The evolution of cooperation in large populations
is usually analyzed in systems based on public goods games \cite{Hardin1968}. 
For each cooperation a cost $c_i>0$ 
depending on the tag of the player is incurred which 
results in a benef\hspace*{-0.001mm}it $b>c_i$ for the 
interaction partner. For simplicity, we restrict ourselves to
two groups of agents only, red and blue. In every group there
are two kinds of players. Intolerant players support only others
 with the same tag ($T=0$). 
Tolerant players ($T=1$) support any other player, regardless of his
group membership. 
The payoff of every player depends on the strategies
of his interaction partners.
We introduce $n^{r}_1$, $n^{r}_0$, $n^{b}_1$ and
$n^{b}_0$ for the number of interaction partners that are tolerant red,
intolerant red, tolerant blue and intolerant blue, respectively.
 The payoff can be written as
\begin{eqnarray}
\Pi(x)\!=
\!\!\left\lbrace\!\!
\begin{array}{lll}
(b-c_r) (n^r_0+n^r_1 + n^{b}_1) -c_r n^{b}_0 & \mbox{for} & x=({\rm red,1})
\\
(b-c_r) (n^r_0+n^r_1)           +b n^{b}_1 & \mbox{for} & x=({\rm red,0}) 
\\
(b-c_b) (n^b_0+n^b_1 + n^{r}_1) -c_r n^{b}_0 & \mbox{for} & x=({\rm blue,1})
\\
(b-c_b) (n^b_0+n^b_1)            +b n^{r}_1 & \mbox{for} & x=({\rm blue,0})
\end{array}\!\!.\right.
\label{payoffeq}
\end{eqnarray}
In the following we restrict ourselves to $c_r=c_b=c$. 
The system with $c_r \neq c_b$ is
analyzed in Section \ref{asym}. 
In a single interaction the payoff is $b-c$ if
both players cooperate with each other, $-c$ indicating
that the player has been exploited, and $b$ indicating that the player
has exploited his interaction partner. The payoff is zero when
both players refused to cooperate. 
The tolerant strategies are 
dominated by the intolerant strategies, as the payoffs 
of the $T=1$ strategies are never higher than the payoffs of 
the $T=0$ strategies. 
In well mixed systems without spatial structure 
this leads to bistability. One group becomes extinct and the 
other group dominates in the two evolutionary stable
Nash equilibria with intolerant players of one tag only.
An alternating dominance of both groups can be 
generated if there is a drift 
towards more tolerance \cite{Traulsen2003}.

\section{Spatially extended system}

Players are arranged on a two dimensional regular 
cubic lattice with periodic boundary conditions. 
The system size is $N=L \times L$.
Each 
player interacts with his eight nearest neighbors 
(Moore-neighborhood), i.e.\ $n_0^r+n_1^r+n_0^b+n_1^b=8$. 
After interacting with all the neighbors,
the players update their strategy synchronously.
This corresponds to discrete, non-overlapping generations.
Strategies are updated
due to the deterministic ``best takes over'' 
rule \cite{Hauert2002a}; 
i.e. $i$ switches to the strategy among its nearest neighbors
that reached the highest payoff;
\begin{equation}
s_i^{t+1}=s_j^t
\;\;\;\;\;  
\mbox{where}
\;\;\;\;\;
j=\mathop{\rm argmax}_{j \in {\rm NN}(i)} \Pi_j^t.
\end{equation}
If several nearest neighbors with different strategies have 
the same success, players keep as much 
of their identity as possible. Choosing between switching
tag or tolerance, players will switch their tolerance. This ensures that
the update rule remains deterministic.
However, these additional rules apply only in very rare cases.
Note that the new strategy of a player depends on the strategies
in his $5 \times 5$ neighborhood, as the payoffs in his $3 \times 3$ neighborhood are
involved. Self interactions can be neglected in our case.
Hence, the game can be described as a 
deterministic cellular automaton with $4^{24}$
update rules. This is in contrast to the usual prisoner's dilemma,
where  ``only'' $2^{24}$ update rules are necessary \cite{Nowak1992}. 
A modif\hspace*{0.001mm}ication of the cooperation cost leads to a
modif\hspace*{0.001mm}ication
of the update rules, see Appendix \ref{coopcost} for details. 
The extension of the usual prisoner's dilemma
to four strategies complicates the application of many
tools for spatial games, as pair approximation 
\cite{Szabo1996,Szabo1998,Szabo2002}, fundamental clusters 
\cite{Killingback1999,Hauert2001}, or 
mapping to Ising models \cite{Herz1994}.

\subsection{Segregation in the basic system}

\begin{figure}[htbp]
\begin{center}
{\includegraphics[totalheight=7cm]{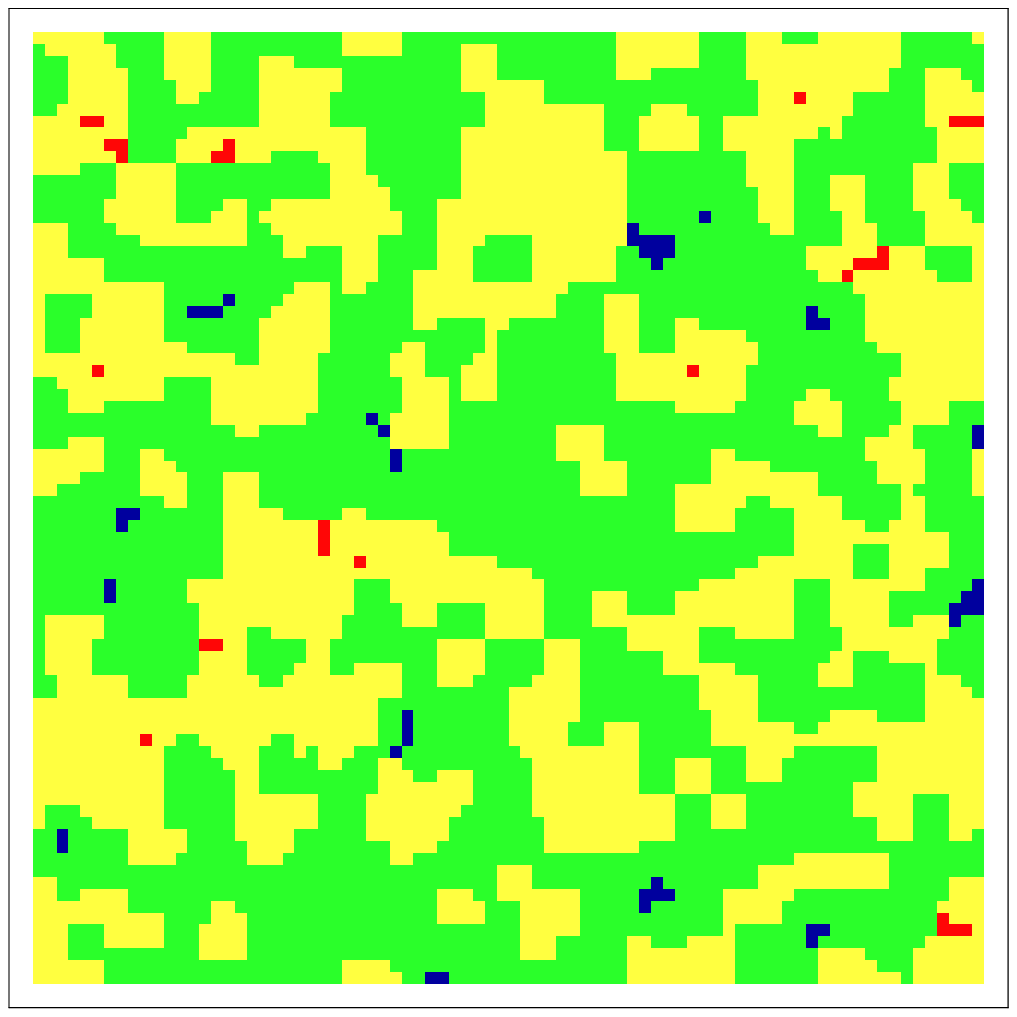}}
{\includegraphics[totalheight=0.5cm]{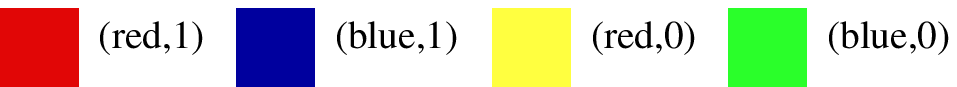}}
\caption{(Color online) Basic system without mutations.  Only 10 
generations after a random initialization the system reaches
a static state. The tolerant players 
can only survive if they have suff\hspace*{0.001mm}icient support
from their neighbors ($c=0.3, b=1.0$, $L=80$).}
\label{nomut}
\end{center}
\end{figure}

As the tolerant strategies can easily be 
exploited by intolerant players from the 
other group, they can only survive with
suff\hspace*{0.001mm}icient support from surrounding
players, cf.\ Fig.~\ref{nomut}. Hence, the majority 
of players will be intolerant when the system
reaches a static state. 
A $3 \times 3$ cluster of intolerant agents
can always survive, as the player in the center has the highest 
possible payoff in his neighborhood.
In general, a tolerant player not interacting with intolerant players 
of the other tag can always survive. If he interacts with such players,
the cooperation cost determines which kind of
clusters are stable.

As expected \cite{Sigmund2001}, the system shows a strong segregation.
Segregation between different agents in cellular automata
has already been observed in the seminal paper of 
Schelling \cite{Schelling1971}. However, in our 
case segregation is not directly 
based on observable traits of others, but on 
mutual support. 
Most of the players are intolerant.
Players that support others across the
segregation borders are always exploited, they
cannot survive if the cooperation cost is too high.
This is consistent with the mean f\hspace*{0.001mm}ield theory
\cite{Traulsen2003}, where only f\hspace*{0.001mm}ixed points 
with intolerant players of one tag are stable.

\begin{figure}[htbp]
\begin{center}
\includegraphics[totalheight=7cm]{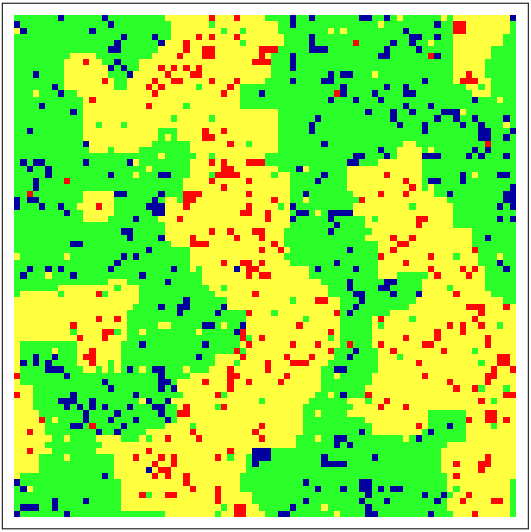}
\caption{(Color online) Basic system with mutations. With $2 \%$
probability, the tag and the tolerance are modified independently.   
As in the system without mutations, the different tags segregate in space. 
100 generations after a random initialization, the majority of small 
clusters seen in the system without mutations, cf.\ Fig.~\ref{nomut}, 
has vanished due to the destabilizing effect
of the mutations. This leads to a longer correlation length, see text.
Colors are as in Fig.~\ref{nomut} 
($c=0.3, b=1.0$, $L=80$).}
\label{mut}
\end{center}
\end{figure}

The situation is slightly different if stochastic 
mutations are included, as the system no longer becomes
static. Tolerance mutations increase the fraction 
of the tolerant agents, as there is no equilibrium between
tolerant and intolerant agents. The tolerance mutations 
lead towards such an equilibrium, while the population dynamics
works against this equilibrium. Mutations of the tags can 
also destabilize clusters, as they introduce new agents 
into an environment that cannot produce such agents by the
population dynamics. This leads to the dissapearance of small
clusters, cf.\ Fig.~\ref{mut}. 

The degree of segregation can be 
quantif\hspace*{0.001mm}ied utilizing the ``spatial dissimilarity index''
$D$ \cite{Duncan1955} def\hspace*{0.001mm}ined as
\begin{equation}
D = \frac{1}{2} \sum_j \left| \frac{N_r^j}{N_r} -\frac{N_b^j}{N_b}\right|,
\end{equation} 
where $N_r^i$ $(N_b^i)$ is the number of red (blue) agents in subregion $i$
and $N_r$ $(N_b)$ is the total number of red (blue) agents.
Choosing a $3 \times 3$ neighborhood as subregion we 
f\hspace*{0.001mm}ind $D=0.715 \pm 0.001$ ($c=0.3$, $N=1000$) 
indicating a strong degree of segregation, 
compared to 
$D=\frac{1}{9} \frac{1}{512}\sum_{j=0}^9 {\footnotesize \left(\begin{array}{c} 9 \\i \end{array} \right)} |2j-9|=\frac{35}{128}\approx0.273$ 
for  a random population. 
$D$ decreases less than $5 \%$ when $c$ is increased
($D=0.729\pm0.001$ for $c=0.05$ and $D=0.707\pm0.001$ for $c=0.95$). 

Another possible measure for the segregation is 
the correlation length $ \lambda$. For simplicity, the correlations
have only been computed for one direction. 
The probability that an agent in the distance 
of $d$ has the same color decays as $ p \propto e^{-d/\lambda}$.  
For $c=0.3$ we f\hspace*{0.001mm}ind a correlation length of 
$\lambda= 5.85 \pm0.02$. The correlation length
decreases slightly with increasing $c$
($\lambda=6.25 \pm 0.02$ for $c=0.05$ and 
$\lambda=5.65 \pm 0.02$ for $c=0.95$). 
As discussed above, mutations lead to the elimination of small clusters. 
Consequently, the correlation length is increased
by mutations. After 100 time steps we find $\lambda = 8.95 \pm 0.03$
($c=0.3$, tags and tolerances are mutated with probability $2 \%$), 
which is significantly higher than the correlation length in the system 
without mutations. 
The segregation properties are not altered if an 
asynchronous update is applied instead.

Overall, the segregation properties and correlation length are 
governed by the length def\hspace*{0.001mm}ined by the size of the neighborhood window.
They are only marginally influenced by the cooperation cost $c$.

\subsection{Emergence of spiral waves from a local conversion mechanism}
\label{biasedconv}

The well mixed system was analyzed rigorously under
the influence of biased conversions towards tolerance
\cite{Traulsen2003}. These biased conversions show an alternating
dominance of both groups in the mixed system.

Let us now introduce a local conversion mechanism
that promotes tolerance in a similar way. 
We assume that an agent in a neighborhood
consisting only of players of his own tag becomes tolerant,
if he did not switch his strategy in the same time step due to selection.
As only intolerant players utilize the ability to distinguish 
between tags this could be motivated by assuming some costs 
for this cognition system.
These conversions lead to a rock-paper-scissors-like cycle
with four strategies: If the neighbors are red, red agents become tolerant.
In a red tolerant neighborhood intolerant blue agents have the highest payoff.
If these dominate the neighborhood, the blue players should switch to
the tolerant strategy. Finally, in such a neighborhood the intolerant red 
agents gain the highest payoff. 
This in contrast to \cite{Szabo2004a}, where 
cyclic dominance is explicitly
included in a system with asynchronous update.

\begin{figure}[htbp]
\begin{center}
{\includegraphics[totalheight=7cm]{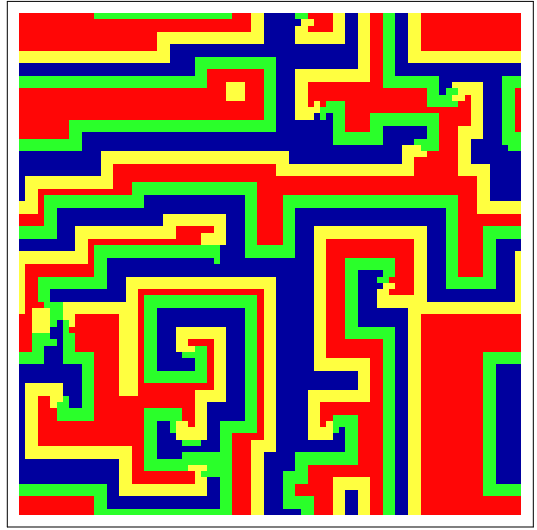}}
\caption{(Color online) System with local conversions towards tolerance.
Colors are as in Fig.~\ref{nomut}.
Agents become tolerant if their new
neighborhood has the same color, this 
leads to a rock-paper-scissors-like cycle. 
Spirals that generate traveling waves appear
($c=0.3, b=1.0$, $L=80$).}
\label{biasedconvpic}
\end{center}
\end{figure}

This cycle leads to the emergence of rotating 
spirals. The arms of these spirals are 
travelling waves, as in the game 
of rock-paper-scissors \cite{Frean2001, Kerr2002}
or in public goods games with volunteering \cite{Szabo2002,Hauert2002b}. 
The front of such a wave 
consists of intolerant agents, these are 
followed by tolerant agents of the 
same tag, cf.\ Fig.~\ref{biasedconvpic}. These players 
can be exploited by intolerant players with different tag,
hence a new front with a different tag can invade.

In the case of asynchronous update spirals are replaced
by larger structures moving through the system. However,
the mechanism for the movement of these structures is the same
as for the spiral waves. 

\label{charges}
To locate the spiral tips quantitatively,
the strategies $(r,1), (r,0), (b,1), (b,0)$
are associated with discrete indices 
$0,1,2,3,$ respectively.
Interpreting those as four possible angles of
a two-dimensional vector f\hspace*{0.001mm}ield, the
curl can be calculated from a counterclockwise 
Stokes path on a $2\times 2$ block.
For a continuous f\hspace*{0.001mm}ield of phases $\phi$ the
topological charge 
of a closed path $\Gamma$ 
is def\hspace*{0.001mm}ined by
\begin{eqnarray}
q = \frac{1}{2\pi} \oint_\Gamma \vec{\nabla} \phi \cdot {\rm d}\vec{r}.
\end{eqnarray}
In our case, both space and phase are discrete; the phase 
is measured in units of $\pi/2$.
Along the Stokes path we compute the phase differences
\begin{eqnarray}
\Delta \phi_1 & =  & x_{i+1,j}  - x_{i,j} \nonumber \\
\Delta \phi_2 & =  & x_{i+1,j+1} - x_{i+1,j}\nonumber \\
\Delta \phi_3 & =  & x_{i,j+1}  -x_{i+1,j+1}\nonumber \\
\Delta \phi_4 & =  & x_{i,j}  - x_{i,j+1}
\end{eqnarray}.
In the discrete case a phase difference of two
steps, or angle $\pi$, may occur,
and consistently can be interpreted as
a zero contribution to the Stokes integral
(leading to the possibility of half-valued
partial charges as discussed below).
Thus the phase differences are mapped on differences
$\Delta q_i$
according to Table \ref{mapping}.
\begin{table}[htbp]
\caption{\label{mapping}
Mapping of phase differences $\Delta \phi_i$ to charge differences $\Delta q_i$.}
\begin{ruledtabular}
\begin{tabular}{ccccccccc}
$\Delta \phi_i$	& $-3$ 		& $-2$ & $-1$ 		& $0$ & $+1$ 		& $+2$ & $+3$   \\ 
$\Delta q_i$   	& $+\frac{1}{4}$	& $0$  &  $-\frac{1}{4}$	&  $0$&  $+\frac{1}{4}$	&  $0$&  $-\frac{1}{4}$\\  
\end{tabular}. 
\end{ruledtabular}
\end{table}
The topological charge
is given by $q=
\Delta q_1
+\Delta q_2
+\Delta q_3  
+\Delta q_4$.
A typical spiral tip consists of two equal 
topological charges $q=\pm \frac{1}{2}$
in nearby positions.
In the stationary regime, the generic case 
is a pairing of two spirals with different
chiralities, i.e.\ different topological charges $q=\pm 1$,
near each other. 
For completeness, it should be noted that the resulting
curl f\hspace*{0.001mm}ield is def\hspace*{0.001mm}ined on the dual lattice shifted
from the original one by a vector $(\frac{1}{2},\frac{1}{2})$.

\begin{figure}[htbp]
\begin{center}
\psfrag{0.00}{0.00}
\psfrag{0.01}{0.01}
\psfrag{0.02}{0.02}
\psfrag{0}{0}
\psfrag{100}{100}
\psfrag{200}{200}
\psfrag{300}{300}
\psfrag{400}{400}
\psfrag{500}{500}
\psfrag{t}{t}
\psfrag{q}[][][1][180]{$\langle \rho \rangle$}
\psfrag{a}{a}
\psfrag{b}{(b)}
\psfrag{c}{(c)}
{\includegraphics[totalheight=8cm]{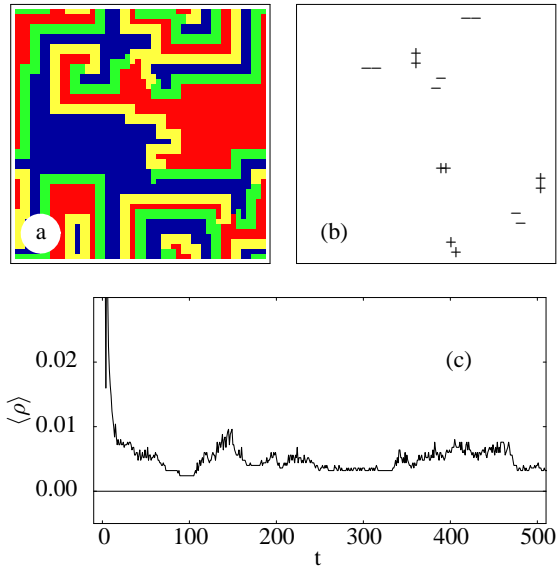}}
\caption{(Color online) Topological charges: Colors are as in Fig.~\ref{nomut}.
(b) shows the spatial distribution of charges for
the strategy distribution shown in (a). The sign $+$~(-) indicates 
a topological charge of $q=+\frac{1}{2}$ ($q=-\frac{1}{2}$).
Full charges are not stable and disappear immediately
after their generation. 
(c) shows a typical time development of the average 
charge density $\langle \rho \rangle$. Initially, 
$\langle \rho \rangle$ drops rapidly. As topological 
charges are generated and annihilated, the charge
density fluctates until the system reaches a stationary state  (L=50).}
\label{chargefig}
\end{center}
\end{figure}

A comparison between the strategy distribution and 
the corresponding charge distribution is shown in
Fig.~\ref{chargefig}.

For a random initialization we f\hspace*{0.001mm}ind an 
topological charge density of 
$\langle \rho \rangle = 0.219 \pm 0.003$,
which is consistent with the theoretical 
average value for independent topological charges 
$\langle \rho \rangle = \frac{7}{32} = 0.21875$.
The spatial game dynamics quickly reduces the initial topological 
charge density. However, 
topological charges are generated and annihilated 
in a irregular manner until the system reaches a
stationary state, see Fig.~\ref{chargefig}.

\section{Asymmetric spatial system}
\label{asym}

\begin{figure}[htbp]
\begin{center}
\psfrag{0.0}{0.0}
\psfrag{0.2}{0.2}
\psfrag{0.4}{0.4}
\psfrag{0.5}{0.5}
\psfrag{0.6}{0.6}
\psfrag{0.8}{0.8}
\psfrag{1.0}{1.0}
\psfrag{Ar}{$A_r$}
\psfrag{Br}{$B_r$}
\psfrag{Cr}{$C_r$}
\psfrag{Ab}{$A_b$}
\psfrag{Bb}{$B_b$}
\psfrag{Cb}{$C_b$}
\psfrag{Dr}{$D_r$}
\psfrag{Db}{$D_b$}
\psfrag{Es}{$E$}
\psfrag{costred}{$c_r/b$}
\psfrag{costblue}[][][1][180]{$c_b/b$}
{\includegraphics[totalheight=8cm]{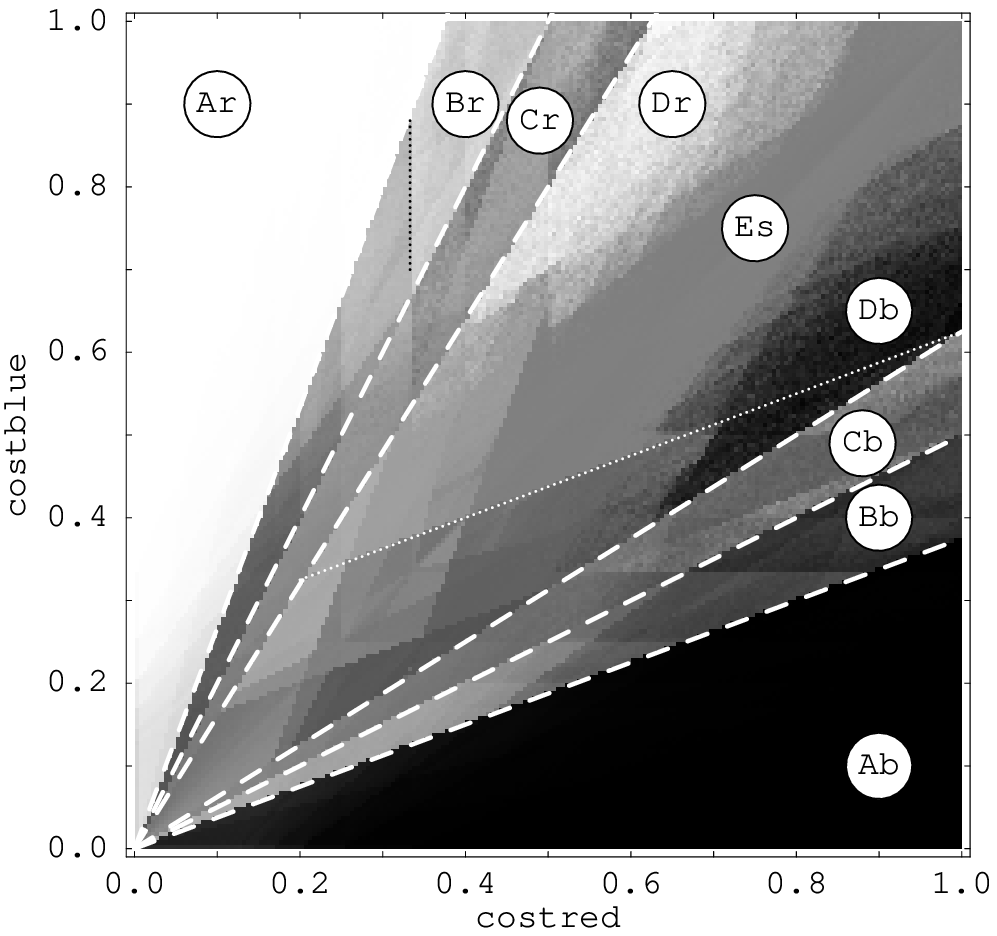}}
{\includegraphics[totalheight=0.6cm]{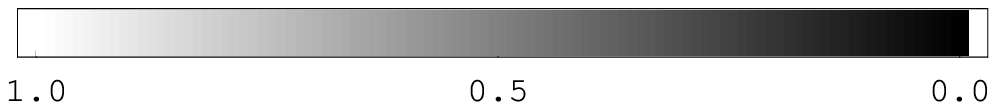}}
\caption{Asymmetric system: Fraction of red agents (encoded in a grayscale) in dependence 
of the cooperation costs $c_b$ and $c_r$.
In region ($A_r$) the population is dominated by red agents, for small $c_b$
intolerant blue agents can survive. In region ($B_r$) intolerant blue players
form channels in a sea of tolerant red agents. In ($D_r$)
red tolerant agents dominate again.
Region ($E$) shows spiral waves as the symmetric game, 
cf.\ Fig.~\ref{biasedconvpic}. Region ($C_r$) shows coexistence
of spiral waves from ($E$) and channels from ($B_r$). 
For $c_b<c_r$ the dynamics is the same with the role of red and blue agents exchanged. 
Structures inside the regions are determined by changes
of the update rule. The borders of these structures, 
e.g.\ the dotted lines in regions ($D_b$)/($E$) and ($B_r$),
are given by linear equations $c_b= a+ b c_r$, see text for details
($L=100$, averages over $50$ time steps and $100$ independent realizations).}
\label{phasediagramm}
\end{center}
\end{figure}

It seems natural to assume that the two
different groups can have two different costs
of cooperation.
For different costs $c_r$ and $c_b$ in Eq.\ (\ref{payoffeq})
we f\hspace*{0.001mm}ind several distinct dynamical regimes,
as dominance of red players, spiral waves, etc.,
cf.\ Fig.~\ref{phasediagramm} for details.

\begin{figure}[htbp]
\begin{center}
\psfrag{0.00}{0.00}
\psfrag{0.02}{0.02}
\psfrag{0.04}{0.04}
\psfrag{0.0}{0.0}
\psfrag{0.2}{0.2}
\psfrag{0.4}{0.4}
\psfrag{0.5}{0.5}
\psfrag{0.6}{0.6}
\psfrag{0.8}{0.8}
\psfrag{1.0}{1.0}
\psfrag{Ar}{$A_r$}
\psfrag{Br}{$B_r$}
\psfrag{Cr}{$C_r$}
\psfrag{Ab}{$A_b$}
\psfrag{Bb}{$B_b$}
\psfrag{Cb}{$C_b$}
\psfrag{Dr}{$D_r$}
\psfrag{Db}{$D_b$}
\psfrag{Es}{$E$}
\psfrag{costred}{$c_r/b$}
\psfrag{costblue}[][][1][180]{$c_b/b$}
{\includegraphics[totalheight=8cm]{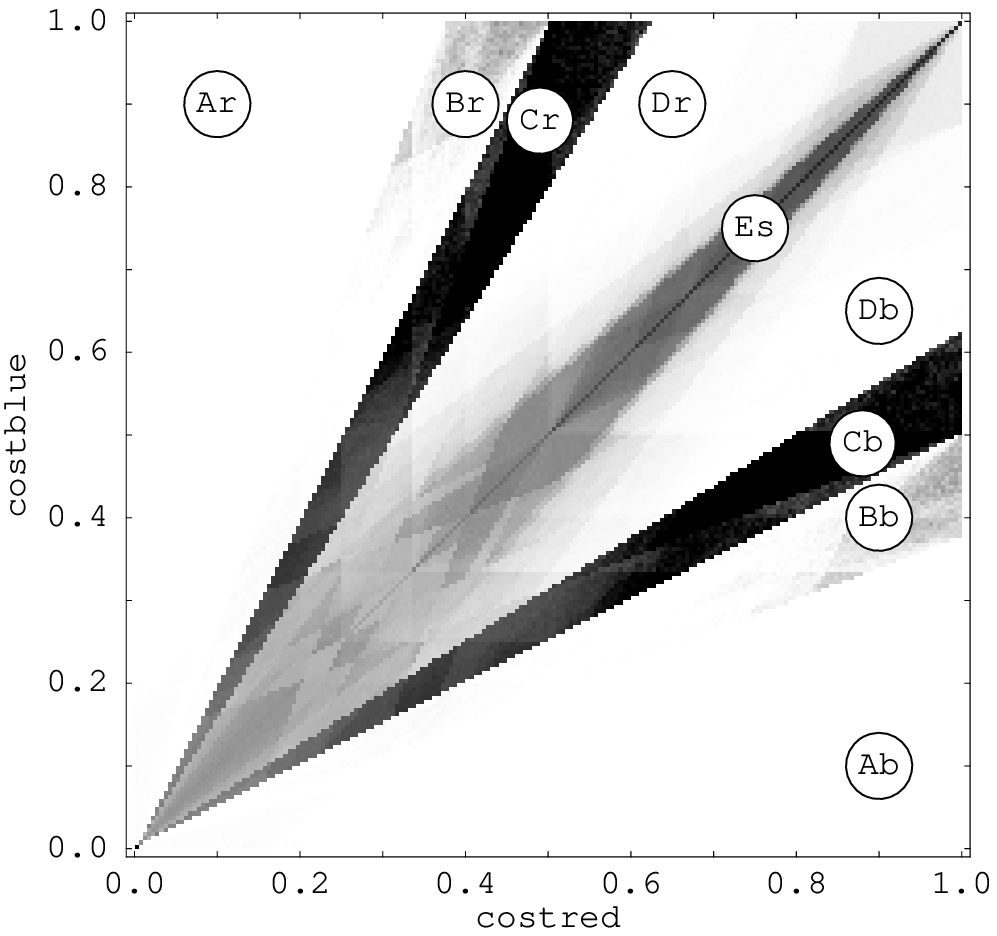}}
{\includegraphics[totalheight=0.6cm]{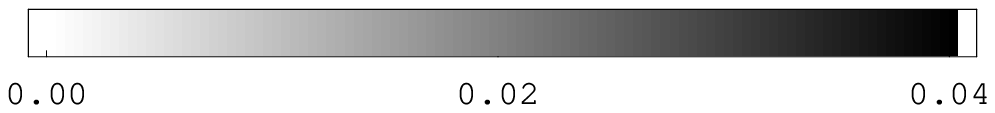}}
\caption{Asymmetric system: 
Average charge density in dependence 
of the cooperation costs $c_b$ and $c_r$.
The regions are the same as in Fig.\ \ref{phasediagramm}. 
Topological charges can
only be found when travelling waves are present, i.e.\ in
regions ($C$) and ($E$). The absence of topological 
charges corresponds to the dominance of one group
in the limit $t \to \infty$.
The highest charge densities are observed
in the ($C$) regions and in ($E$) near the diagonal $c_r=c_b$
($L=100$, averages over $50$ time steps and $100$ independent realizations).}
\label{chargediagram}
\end{center}
\end{figure}

Three different classes of transitions can be 
observed in Fig.\ \ref{phasediagramm}. 
As in the symmetric game,
the update rule is modif\hspace*{0.001mm}ied if one of the cooperation
costs crosses a threshold cost as explained in 
Appendix \ref{coopcost}. Such transitions are 
vertical and horizontal lines in Fig.\ \ref{phasediagramm},
e.g.\ the black dotted line in region ($B_r$) at $c_r=\frac{1}{3}b$.
 Note that the transitions
shown in Fig.~\ref{coopcostpic} (curve L) can be observed
on the diagonal $c_r=c_b$ in Fig.~\ref{phasediagramm}.
A second kind of threshold is determined by $c_r/c_b$. These
thresholds govern the dynamical behavior and divide the phase
plane in Fig.\ \ref{phasediagramm} into the seven distinct regions.
For $\frac{c_b}{c_r}>\frac{8}{3}$ red agents dominate the population, for
$\frac{c_b}{c_r}>\frac{8}{3}$ stationary clusters of intolerant blue agents can survive. 
At $\frac{c_b}{c_r}=2$ travelling waves can appear which suppress
stationary clusters of intolerant blue agents at $\frac{c_b}{c_r}<\frac{8}{5}$.
However, if the cooperation costs are suff\hspace*{0.001mm}iciently high one 
group can take over the whole population after 
a transient period (region ($D_r$) in Fig.\ref{phasediagramm}).   
For  $c_b< c_r$ the roles of red and blue are exchanged.
Finally, we have a third class of transitions which is given by 
linear equations $c_b= a+ b c_r$. Here, the transition threshold is given by a certain 
slope $\frac{\Delta c_b}{\Delta c_r}={\rm const.}$ as for the second kind of transitions.
However, now one of the costs has to exceed a certain threshold
as for the f\hspace*{0.001mm}irst kind of transitions, e.g.\ the white dotted line
in region ($D$) of Fig.\ \ref{phasediagramm} is given by $c_b=\frac{1}{4} b + \frac{3}{8} c_r$. 

It is also posible to describe the phases in the asymmetric 
system by topological charge densities 
introduced in Sec.\ \ref{charges},
cf.\ Fig.\ \ref{chargediagram}. 

\section{Forecast rule}
\label{forecast}

Here, we return to a system where both cooperation costs
are the same.  
Let us assume that the local conversion
rule towards tolerance is based on the {\em new}
strategies of the neighbors and applies also
for players that switched their strategies due to selection. 
Hence, now the update depends on the strategies 
in a $7 \times 7$ neighborhood.
Such a mechanism can be viewed as a 
primitive forecast. Players become tolerant 
if they expect their neighborhood to cooperate
with them in the next generation. Even in this setting
the local conversion rule leads to the emergence of spiral waves. 

It is straightforward to add an 
equivalent mechanism that increases the 
fraction of intolerant agents.
Tolerant agents can become intolerant in order to
protect themselves against exploiters that refuse to cooperate. 
Therefore, we decrease 
the tolerance of an agent if at least two neighbors
will probably exploit him in the next time step.

\subsection{Cooperative fixed point}

\begin{figure}[htbp]
\begin{center}
{\includegraphics[totalheight=7cm]{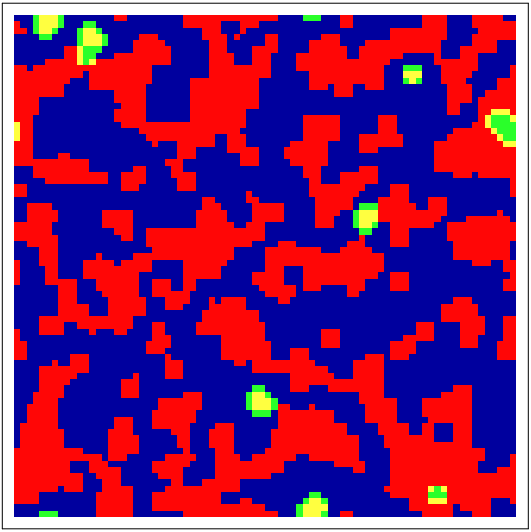}}
\caption{(Color online) System with forecast rule.
Colors are as in Fig.~\ref{nomut}. Agents become
tolerant if their neighborhood has the same 
color and intolerant if at least 2 neighbors will refuse 
to cooperate. This mechanism allows the coexistence of 
red and blue tolerant agents, here the intolerant agents 
have been eliminated by the mechanism, leading to a static state
($c=0.3,b=1.0$, $L=80$). }
\label{bidirconvpic}
\end{center}
\end{figure}

For $c<0.4 b$ we observe cooperation across the segregation borders
for synchronous update. 
For asynchronous updates this fixed point does not become stable.
The forecast rule leads to a stable coexistence of 
red and blue tolerant agents that provide help for everybody 
in their neighborhood, cf.\ Fig.~\ref{bidirconvpic}. 
As a discrimination between different
agents is no longer necessary, this can be seen as a primitive 
mechanism to overcome segregation. However, although
the behavior of all agents is independent of the tags, the different tags 
are still segregated in space. 
The typical correlation lengths are larger than in the system without
conversion mechanisms. For $c=0.3$ we find $\lambda = 8.32 \pm 0.02$.
The spatial dissimilarity index is only slightly higher, 
we observe $D=0.719 \pm 0.005$. 
Surprisingly the mechanism that enables agents to become intolerant
increases the total fraction of tolerant agents, 
as it helps to stabilize tolerant domains. It is interesting that a
mechanism that increases intolerance helps to eliminate intolerance.
However, the mechanism bears some resemblance to the ``tit-for-tat'' strategy in the
iterated prisoner's dilemma \cite{Axelrod1984}, which punishes
others for not cooperating, but can also forgive
defectors reestablishing cooperation. 
One can even
observe different stationary structures that change 
periodically in which intolerant agents survive, these
resemble the ``blinkers'' in the ``game of life'' 
\cite{Gardner1970}. 
For $c > 0.4 b$ the system reaches a 
stationary state only in very rare cases. However,
parts of the system are still dominated by tolerant agents. 
In the case of $c>0.5 b$ this is not longer the case,
here intolerant agents are found in the whole system.

Note that such a forecast rule cannot stabilize the cooperative
f\hspace*{0.001mm}ixed point in the spatial prisoner's dilemma \cite{Nowak1992},
as defectors have always a higher payoff than neighboring cooperators. 

\subsection{Feedback Stabilization} 

For cooperation costs $c>0.5 b$  the tolerant 
f\hspace*{0.001mm}ixed point is unstable. However, we can enforce
cooperative behavior by global feedback on
the cooperation costs. In social systems
this corresponds to adapting taxes with respect
to the state of the society.
Specifying a desired fraction
of tolerant agents $f^{\star}_{\rm tol}$ we update
the cost depending on the current fraction of tolerant agents $f^t_{\rm tol}$ as 
\begin{equation}
c^{t+1}=c^t+\alpha \left(f^{\star}_{\rm tol}-f^{t}_{\rm tol}\right)c^t.
\label{fbeq}
\end{equation}
This mechanism can stabilize points with $f^{\star}_{\rm tol} < 1$
even for $c>0.5 b$. For each $f^{\star}_{\rm tol}$ 
the cost fluctuates around a threshold that is determined
by a change of the update rule, cf.\ Appendix \ref{coopcost}. 
\begin{figure}[htbp]
\begin{center}
\psfrag{0.0}{0.0}
\psfrag{0.5}{0.5}
\psfrag{1.0}{1.0}
\psfrag{0}{0}
\psfrag{100}{100}
\psfrag{200}{200}
\psfrag{300}{300}
\psfrag{400}{400}
\psfrag{500}{500}
\psfrag{t}{t}
\psfrag{cf}[][][1][180]{Cost $c^t$, Tolerant fraction $f^{t}_{\rm tol}$}
{\includegraphics[totalheight=5cm]{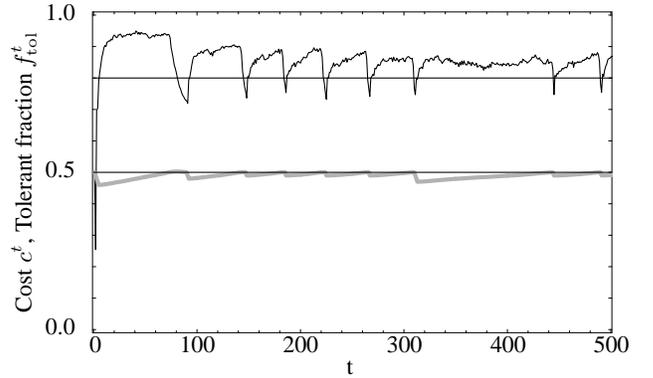}}
\caption{Time evolution of the system with linear feedback.
Cost of cooperation (gray) and fraction of tolerant agents (black).
The fraction of tolerant agents decreases rapidly when the cost
reaches the threshold $c=0.5$  
($f^{\star}_{\rm tol}=0.8$, $\alpha=0.01$, $c^0=0.5$, $N=200$).}
\label{feedbackpic1}
\end{center}
\end{figure}

For the mean f\hspace*{0.001mm}ield theory we have a f\hspace*{0.001mm}ixed point for  $f_{\rm tol}=1$
\cite{Traulsen2003}, which is only stable for very low cooperation costs. 
However, this f\hspace*{0.001mm}ixed point cannot
be stabilized with the linear feedback from Eq. (\ref{fbeq}).

\section{Conclusions}

We have investigated spatial segregation, pattern formation,
and control in a spatial version of a public goods game with 
cooperation based on the similarity between players.
This type of model may establish a useful approach for
a large variety of economical and social systems,
where agents may act not only upon economical considerations,
but also based on similarity or group membership.
Generalizations to more detailed
agents can be performed in a straightforward manner,
yet our four state model already is capable of showing
different phase states from stationary segregation
to complex spatiotemporal behavior.

Particularly, it is interesting to note that a simple
forecast rule can help to overcome segregation
and lead to a stable pattern of cooperating agents,
as regions with limited cooperation at the borders 
between groups are eliminated.However, the different
groups are still segregated in space.

\appendix

\section{Influence of the cost of cooperation}
\label{coopcost} 

\begin{figure}[htbp]
\begin{center}
\psfrag{0.0}{0.0}
\psfrag{0.2}{0.2}
\psfrag{0.4}{0.4}
\psfrag{0.6}{0.6}
\psfrag{0.8}{0.8}
\psfrag{1.0}{1.0}
\psfrag{c}{$c$}
\psfrag{Tolerance}[][][1][180]{Tolerant fraction}
\psfrag{Donationrate}[][][1][180]{Donationrate}
\psfrag{N}{N}
\psfrag{S}{S}
\psfrag{F}{F}
\psfrag{L}{L}
\psfrag{a}{(a)}
\psfrag{b}{(b)}
{\includegraphics[totalheight=5cm]{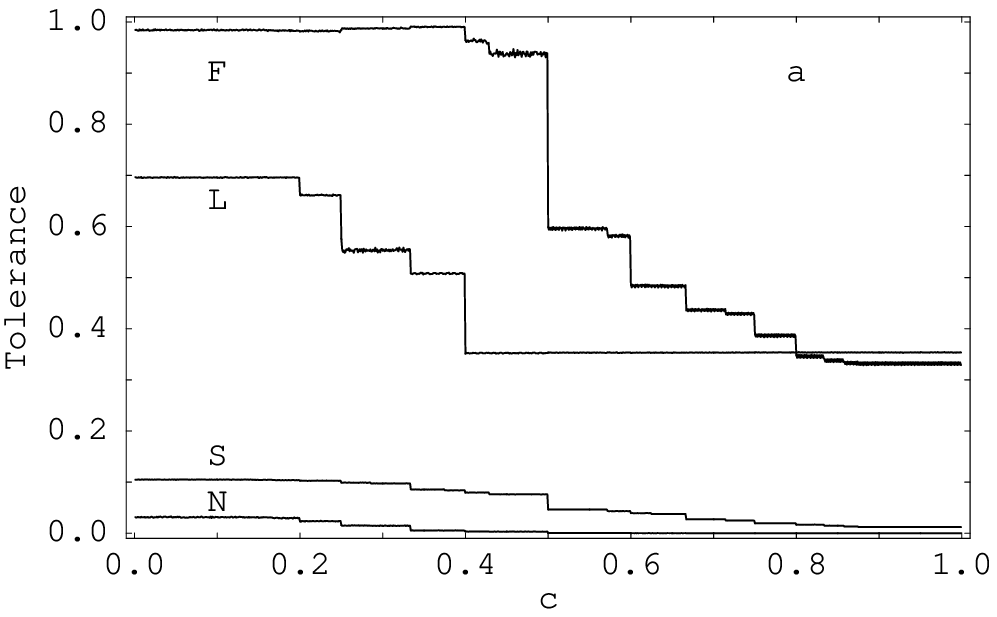}}
{\includegraphics[totalheight=5cm]{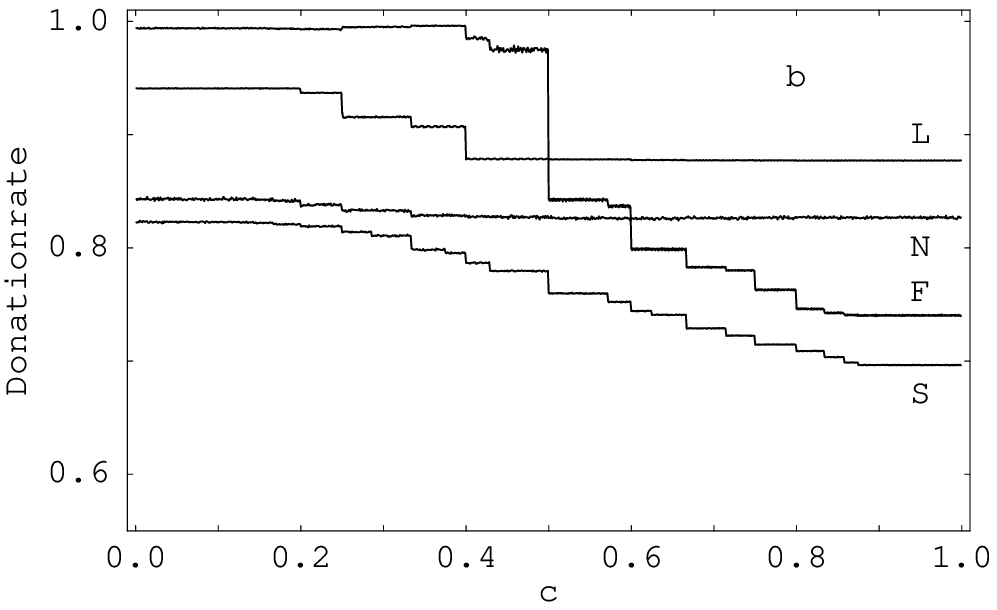}}
\caption{Dependence of different order measures on
the cost of cooperation $c$ for four different update rules. 
(a) shows the fraction of tolerant agents in the system. 
The donation rate, i.e.\ the fraction of interactions in which 
a player cooperated with his co-player, is shown in (b). 
The order measures are shown for the system without 
mutations (N), with stochastic mutations (S), with
the local conversion mechanism described in 
Sec.~\ref{biasedconv} (L), and with the forecast 
rule (F) from Sec.~\ref{forecast}. The sharp steps 
correspond to modif\hspace*{0.001mm}ications of the update rules. 
($b=1.0$, $L=100$, 
spatial averages over $50$ independent realizations and $50$
update steps after a transient period of $50$ update steps).}
\label{coopcostpic}
\end{center}
\end{figure}

Due to the discrete nature of the total payoff,
sharp steps appear when the cost of cooperation 
$c$ varies. As examples, we consider the
dependence of the fraction of tolerant agents $f_{\rm tol}$
and the donationrate, i.e.\ the fraction of cooperative interactions, 
on the cost of cooperation, see Fig.~\ref{coopcostpic}. 
Due to the symmetry between tags,
the fraction of red and blue tolerant agents can be computed
from the fraction of tolerant agents $f_{\rm tol}/2$,
on average. In the same manner the fraction of intolerant 
agents of each group can be computed as $(1-f_{\rm tol})/2$.
The donation rate includes additional information on 
the spatial distribution of the agents. 
The steps that are observed in the order measures correspond to
modifications of the update rule, as described in \cite{Nowak1992}. 
The steps occur at the same positions for all order measures. 
However, the step size is different for the fraction of tolerant agents 
and the donationrate, see Fig.~\ref{coopcostpic}.

\begin{figure}[htbp]
\begin{center}
\psfrag{8b-8c}{8b-8c}
\psfrag{5b-8c}{5b-8c}
\psfrag{5b-5c}{5b-5c}
\psfrag{6b-5c}{6b-5c}
{\includegraphics[totalheight=3cm]{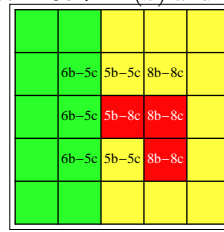}}
\caption{(Color online) Example for a modif\hspace*{0.001mm}ication of the update rule.
Colors are as in Fig.~\ref{nomut}. 
The numbers are the payoffs $\Pi$ for this 
neighborhood. This situtation is stable 
for $c \leq 2/3 b$, as the players with $\Pi = 6b-5c$
that exploit the player in the center do
not have the maximum payoff $\Pi=8b-8c$.}
\label{smallpic}
\end{center}
\end{figure}
Consider a player $x$ with a payoff $\Pi(x)$ and a second
player $y$ with the payoff $\Pi(y)$ and a different strategy. 
If a third player $z$ with a payoff $\Pi(z) < \Pi(x),\Pi(y)$ in
the neighborhood of these two players 
searches for the best strategy,
his update rule changes at $c=c^{\star}$ if the sign of $\Pi(x) - \Pi(y)$
changes at $c=c^{\star}$.
The corresponding values for $c/b$ are given by
\begin{equation}
X^b \;b -X^c\;c = Y^b \;b-Y^c\;c,
\end{equation}
where $0 \leq X^b\;(Y^b) \leq 8$ is the number of agents that support $x\;(y)$
and $0 \leq X^c\;(Y^c) \leq 8$ is the number of agents that are supported by $x\;(y)$.
For the situation shown in Fig.~\ref{smallpic} we f\hspace*{0.001mm}ind for the center player
$\Pi(z)=5b-8c$, for his left neighbor $\Pi(x) = 6b-5c>\Pi(z)$ and for his right neighbor
$\Pi(y)=8b-8c>\Pi(z)$. Hence, we f\hspace*{0.001mm}ind a 
modif\hspace*{0.001mm}ication of the update rule for
\begin{equation}
c^{\star} = \frac{X^b-Y^b}{X^c-Y^c}b=\frac{2}{3}b.
\end{equation}
For $c>c^{\star}$ the center player will switch tolerance
and group membership.
Other transitions can be found in the same way, 
although the method of fundamental clusters \cite{Hauert2001}
is more complicated due to the high number of possible conf\hspace*{0.001mm}\hspace*{0.001mm}igurations.

\acknowledgments{
We thank H.G.\ Schuster for  raising attention to this topic
and stimulating discussions. 
A.T.\ acknowledges support by the Studienstiftung 
des deutschen Volkes (German National Academic Foundation).}

\bibliographystyle{apsrev}

\end{document}